\documentclass{elsart}

\usepackage{natbib}  
 
\usepackage{graphicx}
 
\usepackage{amssymb}

\begin{document}

\runauthor{Campbell}

\begin{frontmatter} 
\title{Ionospheric corrections via PIM and real-time data} 
\author{R.M. Campbell} 


\address{Netherlands Foundation for Research in Astronomy, 
Postbus 2, 7990~AA~dwingeloo, The Netherlands}
 

\begin{abstract} 

   We describe a method for removing ionospheric effects from single-frequency
radio data a posteriori.  This method is based on a theoretical
climatological model developed by the USAF, which returns $n_e(\vec{r},t)$
along the line of sight to the source.  Together with a model of 
{\bf B}$_\oplus$, ionospheric delay and Faraday rotation values ensue.
If contemporaneous ionospheric data --- GPS TEC observations
or ionosonde profiles --- exist, they can be incorporated to
update the modeled $n_e$.

\end{abstract} 



\end{frontmatter} 

\section{Introduction}
\label{intro} 
\vspace{-.2 in}

   The Earth's ionosphere imparts additional phase/group delay 
($\tau_\mathrm{iono}$),  Faraday rotation
($\psi_\mathrm{iono}$), refraction, and absorption to 
cosmic radio signals propagating through it.
These ionospheric effects arise because of the dispersive index of refraction
for a cold, weak plasma (and the presence of 
Earth's magnetic field in
the case of Faraday rotation).  
For $\nu\ge 10$\,MHz, we can safely ignore absorption (see, e.g., 
\citet{Dav1}, \S7.4.1), and will do so henceforth.
We can then write the collision-free Appleton-Hartree index of refraction:
\begin{equation}
\label{eq1}
\mu_p^2 = \frac{2X(1-X)} {2(1-X) - Y_\perp^2 \pm \, [Y_\perp^4 +
4(1-X)^2Y_\parallel^2\,]^{\frac{1}{2}} },
\end{equation}
where $X\equiv ({e^2}/{4\pi^2\varepsilon_0m_e}) \, {n_e}/{\nu^2}$,
$Y \equiv ({e}/{2\pi m_e}) \, {B}/{\nu}$, and the ``$\perp$" and ``$\parallel$"
subscripts for $Y$ refer to $B_\perp$ and $B_\parallel$, 
the components of the Earth's 
magnetic field
perpendicular and parallel to the direction of propagation.
The ``$\pm$"
shows that there are two characteristic modes, giving rise to 
Faraday rotation for linear polarizations.  Equation (1)
can be expanded to yield $\mu_p$ to the desired order
(except for $\nu$ well less than
50\,MHz at times of solar maximum, first-order expansion will suffice).
Any ionospheric effect
will then be proportional to $\int f(n_e, B_\parallel, B_\perp) \, dl$, where
the integration is
along the propagation path.
To first order, $\tau_\mathrm{iono} \propto
\int n_e dl$ [$\equiv$ TEC], and $\psi_\mathrm{iono} 
\propto \int n_e B_\parallel dl$.

  Because ionospheric effects are dispersive, dual-frequency 
observations are the most straightforward way of correcting for them
(again, to first order).  However, such a tactic is not conducive to
some types of observations (e.g., should a source's spectral
index preclude X-band detection).  This paper will discuss one approach
for removing ionospheric contributions from single-frequency data.
This approach comprises two components:  an ionospheric 
``climatology"
model that derives $n_e(\vec{r},t)$ and the capability to
incorporate contemporaneous ionospheric data, if available, to update
the modeled $n_e$.

\section{PIM:  Parameterized Ionospheric Model} 
\label{pim} 
\vspace{-.2 in}

   PIM is a theoretical model of ionospheric climatology developed at USAF
Phillips Laboratory.  It forms the basis for PRISM,
a global real-time ionospheric model for operational use
at the Air Force Space Forecast Center (see \S 3).  PIM compiles runs of
the more general Phillips Lab Global Theoretical Ionospheric Model (GTIM),
spanning parameter space in local time, latitude, season, solar activity,
geomagnetic activity, and interplanetary magnetic-field direction.
GTIM calculates $n_e$ by
solving ion-balance equations along magnetic
flux tubes, taking into account production, loss, and transport processes.
\citet{Dpim1} describes PIM itself; \citet{A1} and \citet{ADV1} discuss
the physics underlying GTIM.
These references also discuss the operational (and philosophical) differences
between theoretical and empirical climatologies.

\begin{figure}[hbt] 
\centering
\includegraphics[scale=0.55,angle=90]{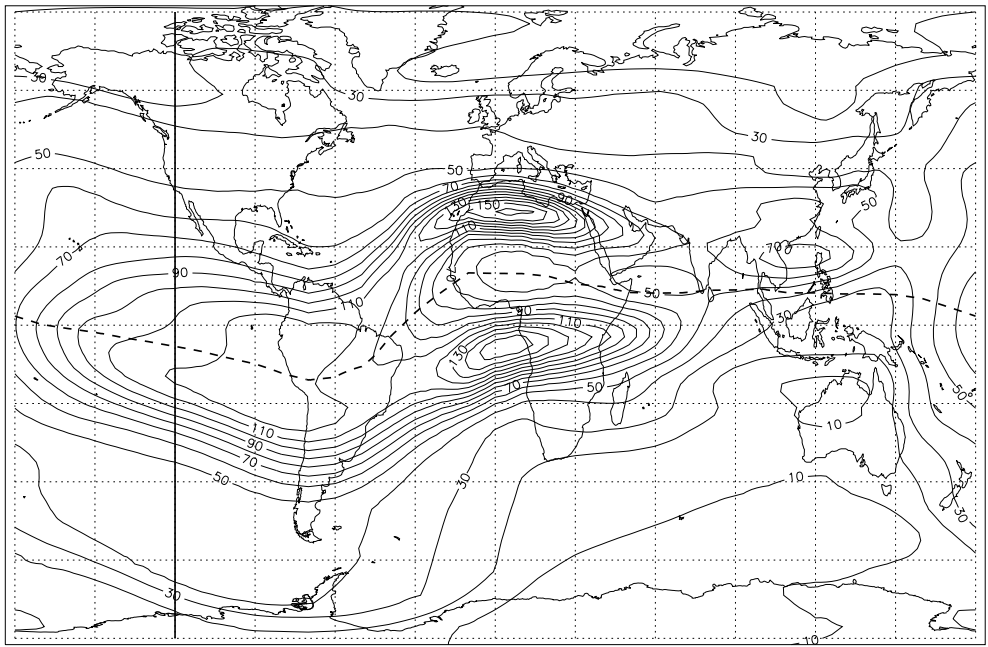}  
\caption{Vertical TEC calculated by PIM at 20UT for the vernal equinox during
solar maximum, in units of TECU.  The longitude grid marks correspond to 2-hr
segments of local mean time; the latitude grid marks fall every $20^\circ$.  
The thick dashed line represents the magnetic equator;
the thick vertical line the longitude of the sun.}
\label{PIMmap} 
\end{figure}

  To illustrate representative ionospheric morphology in PIM,
Fig.~\ref{PIMmap} shows a global map of vertical TEC 
at 20UT around the vernal equinox and solar
maximum.
The contours are labeled in units of TECU ($=10^{16}$\,m$^{-2}$ --- a rule
of thumb:  1\,TECU $\simeq \frac{4}{3\nu}$ cycles of ionospheric delay,
for $\nu$ in GHz).  The thick vertical line represents the longitude of the 
sun, and the thick dashed line is the magnetic equator.  
Roughly speaking, the sun  ``pulls" the entire pattern,
sliding along the magnetic equator, from east to west.
We can see that the predicted horizontal gradients have significant structure
in both E--W and N--S directions, especially for lines of sight
transecting low-latitude regions.
The blobs of anomalously high TEC lying at $\Phi \sim \pm 15^\circ$ 
during the local evening illustrate the interplay
between production, loss, and transport processes in the ionosphere.
Neutral winds in the early afternoon in equatorial regions tend westward.
The differential collisional coupling of the neutrals to electrons and ions
generates an eastward horizontal {\bf E}.  
Acting with the northward {\bf B}$_\oplus$, also predominantly 
horizontal near the equator,
an ${\mathbf E} \times {\mathbf B}$ drift pushes electrons up, to 
regions of lower neutral densities, and hence slower recombination (loss).
These blobs of electrons fall back down along lines of {\bf B}$_\oplus$,
moving away from the magnetic equator and persisting a few hours longer than
they would have otherwise.  This process also causes a diurnal variation in
the height of maximum electron
density ($h_\mathrm{m}F_2$) typically in a range of 250--400\,km.
A general weakness of vertical TEC maps like Fig.~\ref{PIMmap} is that they
mask the fact that vertical $n_e$ profiles can change significantly, in 
their shape and $h_\mathrm{m}F_2$, on time-scales corresponding to diurnal,
seasonal, and solar-cycle factors.

  I am working on a variant of PIM, 
originally motivated by application to
VLBI astrometry (unimaginatively, called PIMVLBI).  
The user needs only to provide the coordinates of the
radio antennas and sources, and the JD/UT of
observations (e.g., input of a Mk\,III/HOPS export file will suffice to pass
all necessary information).
The program automatically looks up the relevant geophysical/solar parameters
from files downloaded from NOAA and GSFC.  Internally, PIM computes $n_e$
at sample points along the propagation path from the source;
PIMVLBI includes
IGRF routines to calculate $B_\parallel$ and
$B_\perp$ at these points. 
The slant  $\tau_\mathrm{iono}$
and $\psi_\mathrm{iono}$ can be
integrated from these values.
This approach takes directly into account
the more realistic horizontal gradients and vertical profiles in PIM,
obviating the need for purely geometric slant factors
about fixed-height sub-ionospheric points, as are ``traditionally"
used for converting
vertical to slant TEC.

\section{Real-Time Data} 
\label{rtd} 
\vspace{-.2 in}

   There are, however, processes whose effects PIM does not model --- global
magnetic storms, traveling ionospheric disturbances, scintillation
(see, e.g., \citet{Dav1}).
For this reason, inclusion of contemporaneous ionospheric observations should
improve the reliability of the modeled $n_e$.  
Much of the data used to constrain
PRISM \citep{A2}
are available only to the military (as is the PRISM code itself).
I have rather incorporated a real-time adjustment (RTA) 
algorithm devised
for range/refraction corrections at an ARPA radar in the Marshall Islands 
\citep{Dkwaj1}.
This RTA distinguishes between two types of real-time data:  integral and 
profile.  

\begin{figure}[hbt] 
\centering
\includegraphics[scale=0.56,angle=90]{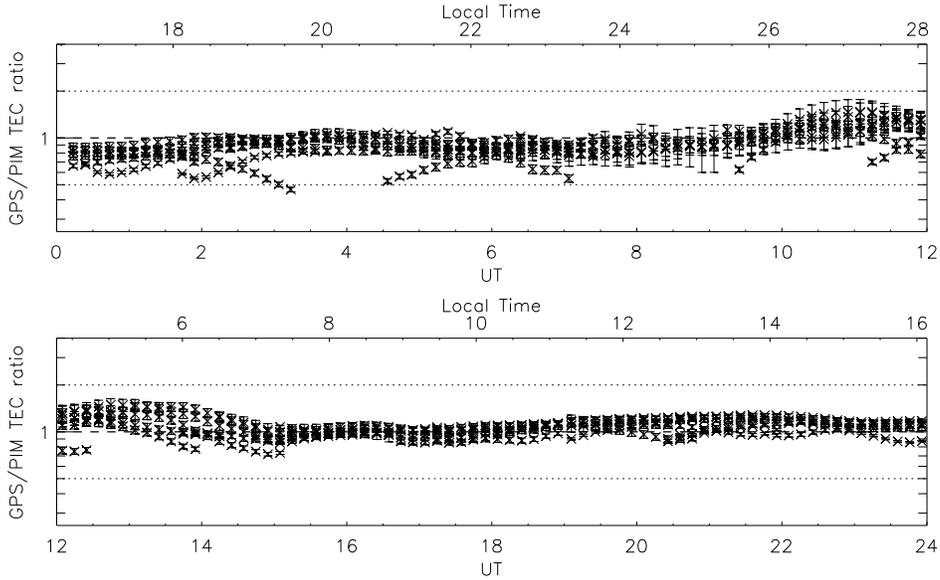}  
\caption{Ratio of GPS-observed to PIM-predicted TEC from 
Goldstone on 9 June 1998.  The $y$-axis is logarithmic, with dotted lines
at ratios of 0.5 and 2.  The local time at Goldstone is labeled on the
upper axes.}
\label{GPScf} 
\end{figure}  

TEC observations from a collocated  GPS receiver 
would be an example of integral data, providing a ratio of GPS/PIM TEC along
the line of sight to each  observed GPS satellite (SV).
In this situation,
one may view PIM as a physics-based mapping function for spatial 
interpolation among the
GPS TEC values to the direction of the source.
Fig.~\ref{GPScf} plots the
GPS/PIM TEC ratios observed from the IGS receiver at Goldstone on 9 June 1998.
For legibility, the 30s-sampled IGS data has been culled to 10-minute sampling.
The error bars stem from the uncertainty in the GPS TEC measurements, 
for which I use
the minimum of the 
running-30-minute standard deviation in the fit of differential
carrier phase and code range for each satellite pass.
Ratios for all observed SVs are plotted regardless of 
their direction.
If PIM has reproduced all spatial gradients perfectly, each such ratio would
equal a single constant.  Scatter within the ratios at a specific time therefore
implies unmodeled structure in $(Az,El)$ space.  We can see that there
are a couple individual SVs diverging from the trend between 18--22\,LT
on that day;
these lie at low elevations off to the south-west, sampling the equatorial 
anomaly.  Except for these cases, the GPS/PIM ratio behaves well, especially so
from 08--16\,LT.

   Ionosondes would be a representative source of
profile data.  The RTA currently uses $h_\mathrm{m}F_2$ and the maximum
density ($N_\mathrm{m}F_2$) from the ionosonde profile.  
Data from ionosondes can neatly complement GPS TEC constraints.  
One problem with the latter is that SVs may not lie along
the lines of sight you would prefer to sample at a given time.
In principle, an ionosonde station could be located where ionospheric
effects along specific lines of
sight from a telescope would be most sensitive to changes in $h_\mathrm{m}F_2$
and $N_\mathrm{m}F_2$.  For example, $h_\mathrm{m}F_2$ rises towards 400\,km
in the equatorial anomaly.  Low elevation southerly lines of sight from 
Europe (as far north as WSRT) intersect this altitude around the African
Mediterranean coast --- just where Fig.~\ref{PIMmap} shows the equatorial
anomaly in the evening.  There are UMass/Lowell digisondes (digital ionosondes)
in southern Spain and Italy \citep{Klob1}; access to their data could 
increase the reliability of PIMVLBI $n_e$ estimates along such directions for
EVN stations, regardless of whether an SV happened to be conveniently 
located.
Other types of profile-data instruments would include incoherent scatter
radars 
and occultations of low-earth
orbiting satellites carrying a GPS receiver.

  There are also other possibilities for expanding the \citet{Dkwaj1} 
RTA.  This was developed to provide real-time corrections for a site
literally in the middle of the ocean; the typical situation for a 
radio-astronomy antenna providing data for later VLBI correlation
is not so severe on both counts.  PIMVLBI already takes advantage of the
relaxation of the real-time requirement
by using data from the complete passes of relevant SVs
when computing GPS TEC values at a specific time (allowing us to
use low-elevation GPS satellites with noise characteristics comparable
to higher-elevation ones when anti-spoofing is on; conversely, 
\citet{Dkwaj1} used
no GPS satellite with $El < 35^\circ$).  However, the 
PIMVLBI RTA still considers
the GPS/PIM ratio ``field" as an independent spatial function each instant.
This leads to situations where, say, an SV yields
a non-unity GPS/PIM ratio at a time just before it sets.  Immediately
afterwards, the information contained in the ratio is lost, even though
the actual ionospheric $n_e$ distribution is not changing so abruptly.
An improved algorithm might be able to incorporate some sort of physically
reasonable temporal smoothing of the spatial ratio-field in response to
such discontinuous changes in the SV sampling.  Further, we may 
also be able to incorporate GPS data from non-collocated receivers nearby.
GPS satellites unfortunately move too slowly to allow proper tomographic
reconstruction of $n_e$ above an array of receivers (as could be done with
the old TRANSIT satellites for meridional arrays), but inclusion of 
PIM and ionosonde
$h_\mathrm{m}F_2$ and $N_\mathrm{m}F_2$ data could perhaps provide sufficient
external constraints to increase the reliability of the slant-to-vertical
and horizontal-translation transformations
required to use non-collocated GPS/PIM ratios in the traditional way.

\section{Concluding Thoughts}
\label{concl} 
\vspace{-.2 in}

   The aeronomy community has made a great deal of progress
in ionospheric modeling
since the mid-70's that has not been thoroughly absorbed by radio
astronomy, and from which we could profit.  The PIM model just described
falls in this category.  Among its advantages, it is freely available and
is upgraded as model refinements proceed.  It calculates $n_e(\vec r)$,
allowing direct integration, to arbitrary order, of effects along the
specific lines of sight from our antennas to our sources.  It can readily
incorporate various types of external contemporaneous data should they be
available, but does not require them should they not.
The next step, once all the code is in place, involves pursuing validation,
especially via the extensive store of IGS GPS data (e.g., more 
``Figures~\ref{GPScf}")
to characterize its strengths and weaknesses as correlated to location,
local time, season, solar cycle, etc.  Such work would complement USAF
PRISM-validation efforts, for which GPS data play a decidedly subsidiary
role to ionosondes and in situ ionospheric data.  There has also been 
preliminary discussion about including ionospheric modeling capability
in AIPS++.  For example (in broadest strokes of AIPS++ parlance), the PIM
databases and geophysical-parameter files could become global data; GPS
TEC measurements from individual stations could become part of their
measurement set; and the user could have the flexibility to
explore various combinations of PIM
and the available real-time data as she sees fit during the course of 
analysis.


\begin{thebibliography}{999}


\bibitem[Anderson, 1993a]{A1}
Anderson, D.N., 1993a, in: Matsumoto, H. (ed), Modern Radio Science 1993,
Oxford University Press, Oxford, p. 159.

\bibitem[Anderson, 1993b]{A2}
Anderson, D.N., 1993b, in: Goodman, J.M. (ed), Proceedings of the 1993
Ionospheric Effects Symposium, NRL, Washington, DC, p. 353.

\bibitem[Anderson et al., 1996]{ADV1}
Anderson, D.N., Decker, D.T., \& Valladares, C.E., 1996, in: Schunk, R.W. (ed),
Solar-Terrestrial Energy Program:  Handbook of Ionospheric Models, NOAA, 
Boulder, p. 133.


\bibitem[Campbell, 1995]{C1}
Campbell, R.M., 1995, Ph.D. Thesis, Harvard University.
%

\bibitem[Daniell et al., 1995]{Dpim1}
Daniell, R.E., Brown, L.D., Anderson, D.N., Fox, M.W., Doherty, P.H.,
Decker, D.T., Sojka, J.J., \& Schunk, R.W., 1995, Radio Sci., 30, 1499.

\bibitem[Daniell et al., 1996]{Dkwaj1}
Daniell, R.E., Millman, G.H., Hunt, S.M., Brown, L.D., Lamicela, J.T.,
\& Sponseller, D.L., 1996, in: Goodman, J.M. (ed), Proceedings of the 1996
Ionospheric Effects Symposium, NRL, Washington, DC, p. 368.

\bibitem[Davies, 1990]{Dav1}
Davies, K., 1990, Ionospheric Radio, Peter Peregrinus, Ltd., London.


\bibitem[Klobuchar, 1997]{Klob1}
Klobuchar, J.A., 1997, Radio Sci., 32, 1943.



\end{thebibliography}
\end{document}